# A Scalable Data Science Platform for Healthcare and Precision Medicine Research

Jacob McPadden MD[1,†], Thomas JS Durant MD[2,3,†], Dustin R Bunch PhD[2], Andreas Coppi PhD[3], Nathan Price[4], Kris Rodgerson[4], Charles J Torre Jr MS[4], William Byron[4], H Patrick Young PhD[3,5], Allen L Hsiao MD[1], Harlan M Krumholz MD SM[3,5,6], Wade L Schulz MD PhD[2,3*]

[1]Yale University School of Medicine, Department of Pediatrics; [2]Yale University School of Medicine, Department of Laboratory Medicine; [3]Yale New Haven Hospital, Center for Outcomes Research and Evaluation; [4]Yale New Haven Health, Information Technology Services; [5]Yale University School of Medicine, Department of Internal Medicine, Section of Cardiovascular Medicine; [6]Yale School of Public Health, Department of Health Policy and Management

[†]These authors contributed equally to the manuscript.
[*]To whom correspondence should be addressed: Department of Laboratory Medicine, 55 Park Street PS345D, New Haven, CT 06511. Tel.: (203) 819-8609. E-mail: wade.schulz@yale.edu

## Abstract

*Objective:* To (1) demonstrate the implementation of a data science platform built on open-source technology within a large, academic healthcare system and (2) describe two computational healthcare applications built on such a platform. *Materials and Methods:* A data science platform based on several open source technologies was deployed to support real-time, big data workloads. Data acquisition workflows for Apache Storm and NiFi were developed in Java and Python to capture patient monitoring and laboratory data for downstream analytics. *Results:* The use of emerging data management approaches along with open-source technologies such as Hadoop can be used to create integrated data lakes to store large, real-time data sets. This infrastructure also provides a robust analytics platform where healthcare and biomedical research data can be analyzed in near real-time for precision medicine and computational healthcare use cases. *Discussion:* The implementation and use of integrated data science platforms offer organizations the opportunity to combine traditional data sets, including data from the electronic health record, with emerging big data sources, such as continuous patient monitoring and real-time laboratory results. These platforms can enable cost-effective and scalable analytics for the information that will be key to the delivery of precision medicine initiatives. *Conclusion:* Organizations that can take advantage of the technical advances found in data science platforms will have the opportunity to provide comprehensive access to healthcare data for computational healthcare and precision medicine research.

## 1. Background and Significance

Healthcare data has seen massive growth over the last several years, with some reports estimating that healthcare data generation increases by 48% annually.[1] In addition, it has been estimated that the intelligent use of big data within the healthcare sector could save over $300 billion.[2] One particular area of medicine that relies heavily on these big data sources is precision medicine, where massive amounts of information are needed to provide precision diagnostics or therapeutics.[3-5] However, efforts to store, manage, and analyze these growing data sets has stretched the limits of traditional healthcare information technology (HIT) systems.[6]

Many definitions of big data exist, with one of the simplest being "any dataset that is too large or complex for traditional hardware or data management tools."[7] In addition to the significant increases in volume, healthcare data are highly complex due to the presence of many data standards and an estimated 80% of information being unstructured.[8] These data can be problematic for traditional enterprise solutions which rely heavily on defined data models prior to indexing, making it difficult to accommodate new data feeds or evolving data structures.[9] To support the informatics needs for the next generation of computational health research, novel approaches to data storage and analysis are necessary.

Fortunately, several applications have emerged that begin to address the key challenges in big data processing, such as distributed data storage and scalable processing capacity.[10] One example is the Hadoop platform, a set of open-source tools designed specifically for these tasks.[11] The goal of these platforms is to create a central repository, called a data lake, which can store raw data in its native format for later search, retrieval, and analysis. However, researchers and clinicians in the healthcare sector looking to leverage modern big data architectures are faced with particular challenges in implementation and little guidance or evidence on the use of these platforms in parallel with production environments.

With the push for population-wide research initiatives such as the Cancer Moonshot[12] and Precision Medicine Initiative (now called All of Us) [3] that will rely on large, complex, interrelated data, institutions need to develop systems that can adequately scale to handle the data influx and provide sufficient capacity for analytic needs. Nevertheless, any new approaches must be attentive to the privacy and reliability requirements associated with healthcare data. Accordingly, we present two use cases that highlight the architecture and implementation of a healthcare data science platform that enables integrated, scalable, secure and private healthcare analytics. These strategies highlight current best practices for data management, system integration, and distributed computing, while maintaining a high level of security and reliability.

### 1.1 Objective

The goal of this paper is to describe how an integrated data lake and analytics platform can be used to provide near real-time access to healthcare and biomedical research data with the ability to conduct computational healthcare research. We describe the implementation of such a platform, which we have named the Baikal Data Science Platform. We highlight the data workflows and use of specialized data storage formats for two common healthcare use cases: continuous patient monitoring and real-time laboratory analytics. These were chosen as they represent large, real-time data sets that are difficult to store in traditional health care data warehouses.

## 2. Materials and Methods

### 2.1 Hardware and Operating Systems

The Hadoop platform was deployed on a thirty-node cluster running CentOS7 (Red Hat, Raleigh, NC, USA). No virtualization layer was used so as to minimize performance overhead. This cluster has a total of 540 processing cores, 13.7 TB memory, and approximately 800 TB of storage distributed among the nodes, with a scalable framework that can be used to add additional capacity. Five additional nodes running CentOS7 were deployed with a distributed total of 60 cores, 320 GB memory, and 5 TB of storage to run Elasticsearch (Elastic, Mountain View, CA, USA). In addition to the core data storage and analysis nodes, four virtual application servers were created: two running CentOS7 and two running Windows Server 2012 R2. A virtual machine running CentOS7 was also deployed as the Ambari management node for the Hadoop cluster.

### 2.2 Software and System Configuration

Hortonworks Data Platform (HDP) version 2.6.0, a commercially-supported Hadoop distribution (Hortonworks, Santa Clara, CA, USA), was installed with three master nodes, three edge nodes, and twenty-four data nodes. Ambari was deployed with Ansible (Red Hat, Raleigh, NC, USA) playbooks and individual Hadoop applications deployed through the Ambari interface. Key software packages, including the Hadoop Distributed File System (HDFS), Zookeeper, Yet Another Resource Negotiator (YARN), Kafka, Storm, and Spark, were installed on these nodes, in high-availability mode when possible. Docker (Docker, San Francisco, CA, USA) was deployed within a Swarm configuration on the three edge nodes. Hortonworks Data Flow (HDF) version 3.0 (Hortonworks, Santa Clara, CA, USA), based on the open-source NiFi software (version 1.2.0.3), was deployed within a Docker container on one edge node. A development environment was also created using Docker Compose to allow for deployment of the key platform applications as a local, virtual cluster (https://github.com/ComputationalHealth/baikal-devenv).

Elasticsearch version 6.2.2 was deployed within Docker containers to five individual nodes. One node was configured as a dedicated master node, two as master-eligible data nodes, and two data nodes. Kibana version 6.2.2 (Elastic, Mountain View, CA, USA) was deployed in a Docker container to one Linux application server. Other software components relevant to the use cases discussed here include version 3.6 of the RabbitMQ software (Pivotal, San Francisco, CA, USA), the Capsule Neuron software (Qualcomm Life, Andover, MA, USA), and the Cloverleaf (Infor, NY, USA) interface and integration engine.

### 2.3 Compression Efficiency Assessment

A Spark application was developed in Scala to compare the storage and analytic efficiency of three file formats: standard comma-separated values (CSV), Avro, and data compressed with the Snappy codec. Data for a one-month monitoring period was extracted for performance testing. Data elements were loaded and then written to HDFS in each file format and the execution time for the read and write efficiency was obtained from the Spark shell interface. This process was repeated on three separate edge nodes and the mean execution time was calculated for each measurement.

## 3. Results

### 3.1. Platform Architecture and Deployment

Core Components

The core software applications within the Baikal platform include features that allow for distributed data storage, message queuing, streaming data processing, distributed computation, and workflow management (Figure 1). Two key systems form the basis of the data storage platform: the Hadoop Distributed File System (HDFS) and Elasticsearch, a NoSQL database platform. Two message queue applications are also used within the data science platform. Kafka is used within the Hadoop environment and RabbitMQ is used on nodes outside of the Hadoop cluster. Streaming data are processed with Storm, a distributed real-time computation system, or HDF, which provides similar features but with a developer-friendly user interface. Distributed batch computation is done with the Spark framework and custom applications. Workflow management and configuration synchronization are done with the Oozie and Zookeeper applications, respectively. Finally, Docker is used for the deployment of custom applications that can be run within the data science platform.

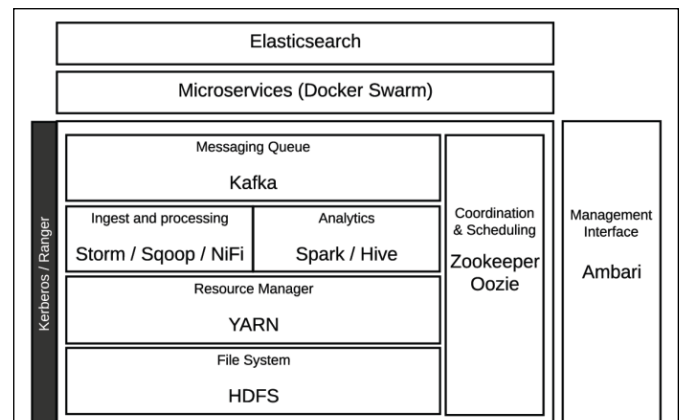

**Figure 1:** *Baikal platform architecture.* Cluster services are monitored, deployed, and provisioned by Ambari management console. Workflow management and configuration synchronization are handled by Zookeeper and Oozie. Data storage frameworks include Hadoop Distributed File System (HDFS) and Elasticsearch. Kafka messaging queues are used for incoming data with subsequent ingest and processing handled by Storm, Sqoop, and NiFi. Analytics can be performed by Spark and Hive. Kerberos and ranger are used to secure cluster applications. Lastly, Docker Swarm is used to deploy custom applications that can be run within the data science platform.

Security

Many big data platforms, including Hadoop, have limited security features enabled by default.[13, 14] For example, no authentication is required to access web service interfaces by default in either Hadoop or Elasticsearch.[14, 15] This lack of default security has led to several data breaches over the last several years.[16] Fortunately, these platforms do allow for configuration of a robust security systems with the use of

Kerberos, Ranger, and Shield.[13, 17] Within the Baikal platform, a dedicated Kerberos realm was deployed for authentication into the cluster. Ranger was deployed to allow for permission-based authorization to resources in the cluster at both the application and data layers.

*3.2. Electron: A Framework for Physiologic Signal Monitoring and Analysis*

Continuous monitoring of patient vital signs has been standard practice in intensive care units and emergency departments. However, these data are rarely kept longer than a few days due to the storage and technical requirements for such large data sets with limited impact for clinical use; however, they may have significant value for future investigation. To support investigators who required access to these physiologic signals, we used the Baikal platform to create Electron, a framework to store and analyze longitudinal physiologic monitoring data. The code for this platform is available within a GitHub repository (https://github.com/ComputationalHealth/electron).

Data Characteristics
Many bedside patient monitors and ventilators are able to transmit their settings and recordings to a central application at specific intervals. Within our institution, these signals are sent at one to five-second intervals, depending on the device, data element, and value. These data elements include active data channels, device/patient metadata, and more intermittent data elements, such as non-invasive blood pressure and alarms. In total, data can be transmitted for up to 892 active and metadata channels for bedside monitors and 45 channels for ventilators. Individual message size varied based on the number of metadata elements, the device being used for monitoring, and frequency of intermittent events. To determine the data storage requirements for such a platform, we collected data for three randomly selected adult and three pediatric patients in the intensive care unit for a 24-hour time frame. A single adult patient in the intensive care unit generated approximately 17.1 MB of data per 24-hour time frame (Table 1). Similarly, a 24-hour monitoring period for pediatric patients averages approximately 12.7 MB in the same timeframe. We similarly assessed the data volume generated by ventilators in our health system, which produced approximately 231.5 MB of data per ventilator per day. When monitoring data from a one-month period was assessed, over six terabytes of raw data from 11 units and a total of 225 beds were collected, often reaching rates of over 400 messages/second. These units were diverse and included intensive care, surgical, emergency department, and short stay beds.

Electron Framework Architecture
The platform to acquire, store, and analyze the continuous monitoring data consists of four key features: data ingestion, data processing and denormalization, compressed storage, and distributed analytics (Figure 2). Our physiologic monitoring infrastructure consists of attached patient monitoring devices that send signals to vendor-supported integration servers (Figure 2A). Data are then transmitted as HL7 messages streamed via a TCP/IP connection to an emissary service that was deployed to accept the incoming message stream and perform the initial conversion of HL7 messages into a custom JSON format (Figure 2B). Date and time information is converted to universal coordinated time (UTC), while all other data are left in their original format. Once processed, messages are forwarded to a secured Kafka message queue, which allows the platform to buffer messages during downstream processor downtime or when under heavy load. The JSON document contains key elements for downstream processing as well as a copy of the original HL7 message to allow for future reprocessing, if needed:

```
{
  "msh_ts":"long",
  "alarm_ts":"long",
  "source":string,
  "unit":"string",
  "text":"string",
  "channel":"string",
  "text":"string",
  "hl7":"string"
}
```

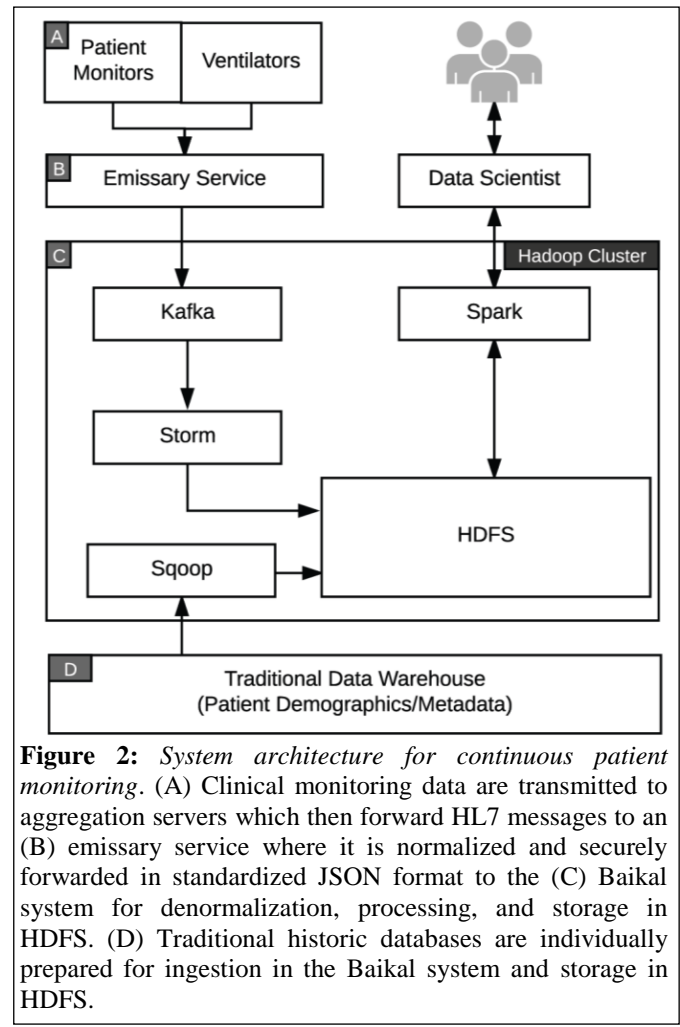

**Figure 2:** *System architecture for continuous patient monitoring.* (A) Clinical monitoring data are transmitted to aggregation servers which then forward HL7 messages to an (B) emissary service where it is normalized and securely forwarded in standardized JSON format to the (C) Baikal system for denormalization, processing, and storage in HDFS. (D) Traditional historic databases are individually prepared for ingestion in the Baikal system and storage in HDFS.

Storage Formats and Data Processing
While storage costs continue to decline, the cost of long-term data storage for large data sets remains burdensome. Specialized data formats and compression can improve the density of data storage, but often come at the cost of increased overhead for read and write throughput. Fortunately, the frequency of access to historic data typically decreases over time which means that slower data access methods would have less impact on overall analytic capacity. Other work has compared the storage and access efficiency for many big data technologies.[18] For this use case, we predicted that the Avro data format with Snappy compression would have an appropriate balance of storage and access efficiency.

| Source | Signal Counts | Storage Size (MB) | Estimated Annual Storage (GB) |
|---|---|---|---|
| *Adult Monitor* | 291,252 (± 84,568) | 17.1 (± 5.0) | 6.2 |
| *Pediatric Monitor* | 223,387 (± 29,543) | 12.7 (± 1.8) | 4.6 |
| *Adult Ventilator* | 3,504,162 (± 236,672) | 231.5 (± 30.6) | 84.5 |

**Table 1:** *Average storage requirements for adult and pediatric patient monitoring, and ventilator monitoring:* Signal counts and storage size represent the metrics for a complete 24-hour per bed monitoring period averaged from three independent samples.

Avro is a semi-structured data serialization format designed for big data storage. In addition to the semi-structured nature of the Avro format, the files are also splittable, which means that the Hadoop platform can separate the file into individual sections which increases the processing efficiency during data analysis.[19] To assess the impact of the Avro format and Snappy compression, we assessed the storage and access efficiency of monitoring data from several different variables over a 30-day period in comma separated raw text, Avro, and Snappy-compressed Avro formats. Data were filtered and the length of time needed to write and read data from three independent nodes in the cluster was recorded. Compared with raw text, Avro-formatted files required approximately 12% more storage space on disk but showed significantly faster data retrieval time (Figure 3). The use of Snappy compression showed significant savings in storage requirements, with an average reduction in file size of 80.5% compared with raw text files. Also of note was the large reduction in time needed to access data stored in Snappy-compressed Avro files.

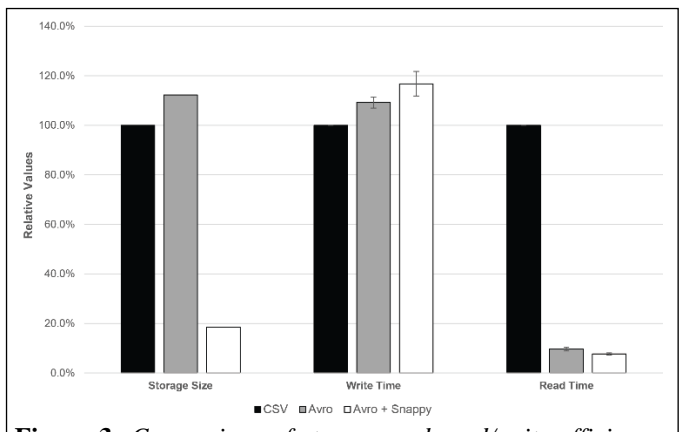

**Figure 3:** *Comparison of storage and read/write efficiency.* Avro increases storage space and write time modestly while significantly reducing read time. The addition of Snappy compression increases write time minimally, while significantly decreasing storage space and maintaining minimal read time. The resulting combination optimizes for single archival write with multiple read usage.

In addition to the large volume, the high velocity of these data required a high-throughput data processing pipeline to convert and store the data efficiently. To achieve this, we developed a custom application built on the Storm platform that allowed for distributed, high-throughput processing. Within the Storm topology, monitor signals were denormalized, converted to the Avro format, compressed with the Snappy codec, and stored in HDFS to allow for future analysis. A separate copy of the data containing the original HL7 message was also stored through a separate Storm bolt in case reprocessing of the raw data became necessary.

Analytics

Much like the particular challenges for the acquisition and storage of big data, specialized needs for the analysis of these data sets also exist. While the raw data are of use for many research and clinical projects, derived variables and predictive analytics are often of interest. For example, computationally-derived features, such as R-R intervals,[20] indices of multiple vital signs,[21] and temporal relations between vital signs have all shown promise as predictive variables.[22] However, generating these features is often computationally intensive when performed at scale on entire patient populations.

Traditional analytic methods and tools are often unable to scale to meet the needs of these analyses. Even in cases where parallelized computation can be used, the resources necessary to develop and validate these custom applications is often prohibitive. To make parallelized computation more accessible, solutions such as MapReduce[11] and Dryad[23] have been created, which provide frameworks that manage the complexity of parallelization. However, these solutions still require significant technical expertise to develop applications that can be deployed to production environments. Within the Baikal platform, we enabled Spark as the primary data analysis tool for batch analysis. Spark is a general data processing framework that can be used to write applications in several common languages, including Java, Scala, Python, and R. A key advantage of this framework is the ability to maintain data for MapReduce operations in memory, rather than needing to read and write each intermediate step to disk. This has been shown to improve the speed of big data processing significantly.[24, 25] We developed several Spark applications that can be used by data analysts to generate features from the physiologic data, such as alarm events, and extract subsets of information for downstream processing, which are available within the GitHub repository.

*3.3. Nucleus: A Platform for Real-Time Laboratory Business Intelligence and Data Visualization*

In addition to novel data sources such as continuous patient monitoring, data science platforms can also offer new approaches for the analysis of more traditional healthcare data sets. Examples include real-time data analysis, predictive analytics, and interactive visualizations. In the era of cost reduction and an increasing demand for clinical laboratory services, laboratorians are facing expectations to optimize laboratory efficiency for the sake of clinical workflows and improve test utilization without compromising quality and safety. Therefore, the clinical laboratory has a particular need for real-time business intelligence (BI) to improve testing efficiency and patient safety.[26] To achieve this, we created a data science platform with BI dashboards to monitor testing within our institution's clinical laboratory (https://github.com/ComputationalHealth/nucleus).

Data Characteristics

Laboratory orders and results are often routed through multiple systems as they transit between the EHR and laboratory instrumentation. This typically includes message integration services and middleware platforms that manage the flow of data between systems created by a number of different vendors. Within our health care system, approximately 40 million individual results are generated annually from six hospitals, 26 satellite locations, and approximately 220 laboratory instruments. A principal challenge for these data is to provide real-time access and visualizations to end users who need actionable insights from these disparate systems. Because of these unique needs, many downstream architectural decisions varied from the continuous monitoring application described.

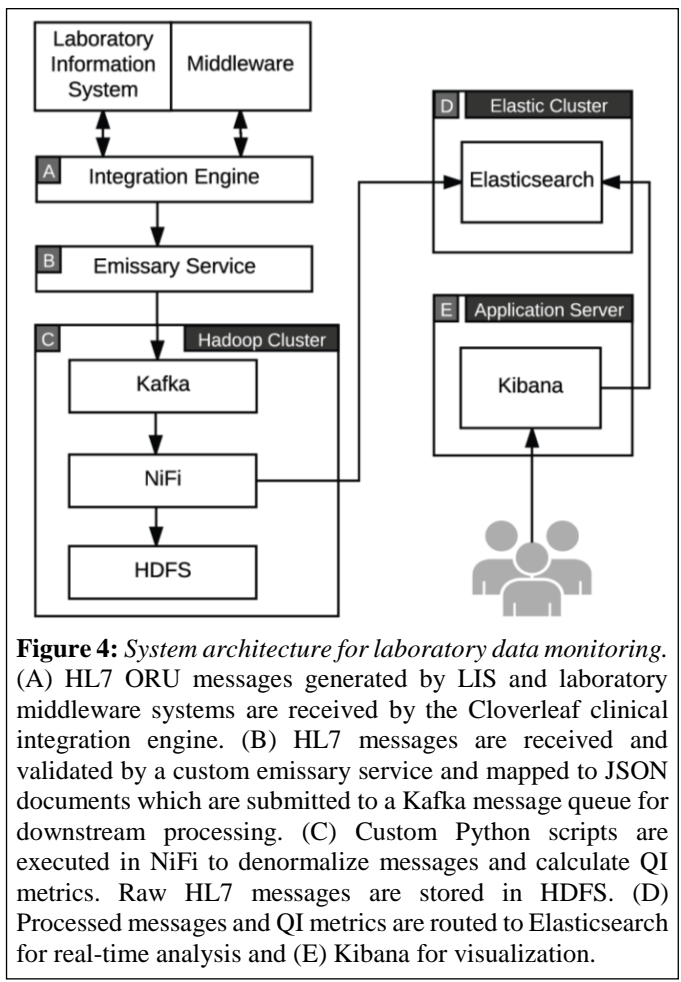

**Figure 4:** *System architecture for laboratory data monitoring.* (A) HL7 ORU messages generated by LIS and laboratory middleware systems are received by the Cloverleaf clinical integration engine. (B) HL7 messages are received and validated by a custom emissary service and mapped to JSON documents which are submitted to a Kafka message queue for downstream processing. (C) Custom Python scripts are executed in NiFi to denormalize messages and calculate QI metrics. Raw HL7 messages are stored in HDFS. (D) Processed messages and QI metrics are routed to Elasticsearch for real-time analysis and (E) Kibana for visualization.

Nucleus Platform Architecture

The initial acquisition of data for this stream is similar to that of the continuous monitoring workflow (Figure 4). Briefly, an emissary service was deployed to receive an HL7 stream of clinical observations and results (ORU) messages from the Cloverleaf integration engine. Each HL7 message was validated and mapped to a JSON document by the emissary service, then forwarded to a secured Kafka message queue. The custom JSON messages contain key parameters that can be used to index and parse results during batch analysis:

```
{
  "msh_ts":long,
  "pt_mrn":"string",
  "order_id":"string",
  "lab_type_code":"string",
  "order_ts":long,
  "hl7":"string"
}
```

Because of the slower message velocity, this data stream was easily processed with the NiFi/HDF software, which is designed for real-time data processing. Custom Python scripts were created to process and denormalize the incoming data stream. Each order and result message was then written to HDFS for permanent storage and batch analysis and also routed to Elasticsearch for real-time analysis and visualization. Additional features that provide key indicators of laboratory efficiency were generated in real time from the HL7 messages with custom Python scripts that are executed within the NiFi workflow. These quality indicators are stored within Elasticsearch and can be used to visualize turn-around time (TAT) for lab results, outstanding orders, and order volumes by patient or laboratory location.

**4. Discussion**

Healthcare information is inherently complex, often has an evolving data structure, and much of the data is not stored in the EHR. Because of this, novel approaches to data management are needed to integrate the many sources of healthcare data. In addition, novel approaches to data analysis such as machine learning require significant computational resources for timely analysis. As the use of big data in healthcare continues to increase, the implementation of robust technical solutions to manage and analyze the data will be important to the success of biomedical big data research.[3]

In this article, we have presented the successful implementation of a data science platform along with two domain-specific applications deployed within this platform. These applications focused on the storage of high volume, real-time data sets that challenge traditional data warehousing strategies due to their volume and velocity. We have also presented the hardware and architectural approaches used to manage these data. While individual components of the platform used here are described in the non-medical literature, this platform combines available technologies to meet the known challenges of big data with needs specific to healthcare, including the security and privacy needs of personal health information.

Often, a single technical solution is unable to address all concerns or needs for a robust data science environment. For example, Hadoop has traditionally been used as a platform for big data storage and batch analysis but had fewer tools available for streaming data and real-time analytics. Because of this, we integrated components designed specifically for the management and visualization of real-time data. This integration allows us to provide efficient batch analytics as well as real-time visualizations, which would be challenging if only a single tool or platform were used. It should be noted, however, that the applications described here are rapidly evolving and significant strides have been made to expand the features of each component, which may add redundancy between applications in the future.

Data science platforms such as Hadoop offer many individual components to address key requirements for data replication, availability, and security at each stage of the data life cycle, from acquisition to analysis. Fully implementing each of these utilities can make data science pipelines complex, but the use of service-oriented architectures affords the ability to update individual applications, scale services, and reuse individual components in multiple workflows. Because of these rapid developments and the diversity of data, careful testing should be done during the implementation of data science workflows to determine the storage and compute capacity required for long-term management of the data being obtained. Similarly, careful attention should be paid to the implementation of built-in security features to ensure that data are not accidentally made available to unauthorized users.[13]

While data science platforms offer significant potential for the rapid analysis of big data, several limitations exist. In particular, the complexity of these platforms often requires substantial technical expertise to use them to their full potential. Multiple software applications are often needed to implement an entire workflow, particularly within the Hadoop environment. While each Hadoop component often provides significant advantages from developing new applications, personnel with expertise are needed to implement these technologies effectively. While many attempts have been made to make the environment fluent with other tools, such as Python, SAS, and R, seamless integration with these tools remains difficult, particularly in secured environments.

Massive resources have been dedicated to big data and data science in other industries; however, the return on investment has not always been realized. Therefore, the ultimate success of these platforms for computational health research will depend on the ability of the biomedical research community to apply big data to translational and clinical research. To fully assess the impact of these systems on healthcare delivery efficiency, additional studies on their implementation and impact will be needed.

## 5. Conclusion

The paucity of literature describing implementation experiences leaves those interested in developing big data environments largely unguided, particularly within the healthcare sector which has unique data and regulatory requirements. Careful attention to the architecture used to create these data science environments will provide an important foundation for future studies that create value from big data sources. As the volume and velocity of healthcare data continue to increase, additional analyses on the management of these data will be required to ensure that the highest quality data is made available to efficient analytic systems.

## Funding and Acknowledgements

This research did not receive any specific grant from funding agencies in the public, commercial, or not-for-profit sectors. We would like to acknowledge Steven Shane and Richard Hurley from Yale-New Haven Health Information Technology Services for their assistance with platform implementation and maintenance. We would also like to thank Guannan Gong for his review of the manuscript.

## Competing Interests

Dr. Krumholz was a recipient of a research grant, through Yale, from Medtronic and the U.S. Food and Drug Administration to develop methods for post-market surveillance of medical devices; is a recipient of research agreements with Medtronic and Johnson & Johnson (Janssen), through Yale, to develop methods of clinical trial data sharing; works under contract with the Centers for Medicare & Medicaid Services to develop and maintain performance measures that are publicly reported; chairs a Cardiac Scientific Advisory Board for UnitedHealth; is a participant/participant representative of the IBM Watson Health Life Sciences Board; is a member of the Advisory Board for Element Science and the Physician Advisory Board for Aetna; and is the founder of Hugo, a personal health information platform.

Dr. Schulz is a consultant for Hugo, a personal health information platform.